# On the utility of cloth facemasks for controlling ejecta during respiratory events

Vivek Kumar,[a] Sravankumar Nallamothu,[a] Sourabh Shrivastava,[a] Harshrajsinh Jadeja,[a] Pravin Nakod,[a] Prem Andrade,[a] Pankaj Doshi,[b] Guruswamy Kumaraswamy[c]

[a] Ansys Software India Pvt. Ltd., Hinjewadi, Phase-1, Pune 411057, Maharashtra, India.
[b] B1-1510, Blue Ridge Township, Hinjewadi, Pune, 411057, Maharashtra, India.
[c] Chemical Engineering Department, Indian Institute of Technology-Bombay, Mumbai 400071, Maharashtra, India.

ABSTRACT

The utility of wearing simple cloth face masks is analyzed using computational fluid dynamics simulations. We simulate the aerodynamic flow through the mask and the spatial spread of droplet ejecta resulting from respiratory events such as coughing or sneezing. Without a mask, a turbulent jet forms, and droplets with a broad size distribution are ejected. Large droplets (greater than about 125 μm in diameter) fall to the ground within about 2 m, while turbulent clouds transport a mist of small aerosolized droplets over significant distances (≈ 5 m), consistent with reported experimental findings. A loosely fitted simple cotton cloth mask (with a pore size ≈ 4 microns) qualitatively changes the propagation of the high velocity jet, and largely eliminates the turbulent cloud downstream of the mask. About 12% of the airflow leaks around the sides of a mask, considering a uniform gap of only 1 mm all around, between the face and the mask. The spread of ejecta is also changed, with most large droplets trapped at the mask surface. We present the viral load in the air and deposited around the person, and show that wearing even a simple cloth mask substantially decreases the extent of spatial spread of virus particles when an infected person coughs or sneezes.

INTRODUCTION

There is consensus[1,2,3,4] that the use of surgical facemasks and N95 respirators help control the transmission of respiratory diseases such as influenza. Therefore, the use of these personal protective equipment (PPE) has been recommended[5] for health care personnel, infected patients and their care givers. To ensure that surgical masks and N95 respirators are available to those at greatest risk of infection, the World Health Organization[6] and national agencies have recommended against their use by the general public. Even so, greatly increased demand during the Covid-19 pandemic has seen a global shortage of such PPE.

Since surgical masks and N95 respirators are not readily available to the public, the use of simple homemade reusable cloth facemasks has been suggested[7,8] as a protective measure, especially due to the possibility[9] of asymptomatic disease transmission. However, there has been considerable controversy[10,11,12,13] over the efficacy of home-made reusable cloth face masks. One report[14] suggests that the use of a homemade face mask would be better than "no protection at all". In South East Asian countries, the practice of using face masks for combating pollution or for personal hygiene is widespread. This has been cited[15] as an important factor in controlling the transmission of Covid-19 infections. However, there are concerns that the use of facemasks may decrease the rigour in following strict physical distancing and handwash hygiene. Further, incorrect use of masks, for example, wearing masks incorrectly or touching the outer surface of masks can result in adverse outcomes. SARS-CoV2 has been found[16] on the surface masks worn by infected patients, and has been shown[17] to remain viable on the surface of surgical masks for several days. Therefore, a detailed understanding of the benefits of wearing homemade cloth masks would be useful in determining policy on their recommended and possible mandatory use.

SARS-CoV2, the virus responsible for the Covid-19 pandemic, infects cells in the upper respiratory system. Transmission of Covid-19 is currently believed[18,19,20] to happen primarily through shedding of virus particles in droplets ejected as infected people speak, cough or sneeze, or through contact with viable infective virus deposited on surfaces. When people cough or sneeze[21] (or even simply talk loudly[22,23]), they eject droplets of mucosal fluid. Large droplets ~O(100 μm) fall due to gravity and, under no wind conditions, are transported over lateral distances of the order of 1 m. However, turbulent flows resulting from violent expulsions during sneezing or coughing suspend finer droplets and transport them over large distances, of the order of 7-8 m.[24,25,26] Therefore, it has been suggested that transmission of infection through fine droplets be investigated.[27,28,29] The effect of surgical masks and N95 respirators on airflows (but not spread of droplet ejecta) during expiratory events has been experimentally imaged.[30] Here, we employ Computational Fluid Dynamics (CFD) simulations to address the influence of home-made face masks on the turbulent clouds that result due to sneezing events, and on the lateral extent of spread of ejecta. Our emphasis is on understanding the effect of face masks in altering the flow field and droplet dispersion due to the respiratory event.

Respiratory events (sneezing or coughing) and the resultant spread of ejecta are modeled as two-phase flow using Ansys Fluent software 2020 R1. The carrier fluid (air) is represented as a continuum phase, and mucosal droplets are represented as the discrete phase. Mucosal droplets are assumed to have properties of water. We model the dynamics of turbulent air jets using time-averaged Navier-Stokes mass and momentum conservation equations. The Renormalization Group (RNG) k-epsilon model is used to model the turbulence, allowing us to span high velocity turbulent flows to lower velocity flows. We employ the enhanced wall treatment model in combination with RNG k-epsilon model to account for the viscous sublayer near the wall surface. A detailed description of the equations and models is given elsewhere,[31] and a brief summary is provided in the supporting information (Section A). For the discrete droplet phase, equations of motion are solved for each droplet to model its trajectory. Droplets can exchange mass (due to vaporization), momentum and energy with the continuous fluid phase. Droplets are convected by the continuous phase and can, in turn, affect the flow of the continuous phase (two way coupling between the fluid and droplet phase). To reduce computational time, half the domain is simulated assuming symmetry in geometry and flow features. Ambient conditions (35°C, 60% relative humidity) representative of summer conditions in India are assumed.

In our simulations, a human face is included in the domain 2 m from the left surface, at a height of 1 m from the ground level, and the mouth is represented as a 2 cm$^2$ opening facing right. Experimentally, it has been observed[24,30,33] that respiratory events result in jets angled towards the ground, with some variability in the angle. We followed previous simulations[32] that model the respiratory event as a jet emanating from the mouth in the horizontal direction. For this jet, a time dependent velocity profile is applied with the peak velocity of 50 m/s at 0.1s. Values for the peak flowrate (= 6 L/s) and the total volume expelled (= 1 L) are obtained from Gupta et al.[33] We do not describe the detailed breakup of the ejecta into droplets. Rather, we prescribe the size distribution of droplets in the ejected spray as a Rosin-Rammler distribution with droplet sizes ranging from 1 to 500 μm. Droplets ejected are not allowed to coalesce or break-up. Based on previous work,[32] we consider that 94% of injected droplet mass can evaporate and the remaining 6% represents non-volatile matter. Following Aliabadi,[32] we inject 6.1 mg of droplets over 10 equal injections with initial droplet positions staggered over 1 cm.

The cloth mask is not tightly fitted around the face, representative of homemade masks. Therefore, we model the mask as covering 22% of the face area around the nose and mouth and model air leaks by considering a uniform gap of 1 mm all around, between the face and the mask. The area of the gap around the mask is about 2% of the mask area. To model[34] the resistance presented by the mask to the flow of air, we consider the mask as an isotropic porous medium, with Darcy and inertial contributions to the pressure drop, given by:

$$S_i = -\left(\frac{\mu}{\alpha}v_i + \sum_{j=1}^{3} C_2 \frac{1}{2}\rho|v|v_j\right) \quad (1)$$

where $S_i$ is the source term in the momentum equation, μ is the fluid viscosity, $v_i$ is the velocity, $\alpha$ is the permeability and $C_2$ is the inertial resistance factor. We consider a mask made of cotton cloth, and obtain $\alpha$ and $C_2$ from a fit to experimental data presented in the literature[35] (details in SI – Section A). To model the permeation of droplet ejecta, we follow experimental reports[36] of the penetration efficiency of fabric. We consider an effective pore size of 4 μm for the cotton cloth and impose a trap condition so that all droplets larger than the pore size are trapped by the mask. More details about the governing equations, computational method and validation can be found in the supporting information (Section A).

The instantaneous flow field 0.1 s after the respiratory event without and with masks, respectively, is shown in figures 1 a, b (close ups: Figures 1 c, e). We represent the flow field using streamlines and present data in the symmetry plane passing through the face. Without a mask (Figure 1 a, c), the respiratory event results in an air jet with a fast moving turbulent core that entrains ambient air and slows down as it propagates from the face. As the surrounding air is entrained, the jet forms a conical shape with a cone angle = $\tan^{-1}$(radius/height) ≈ $\tan^{-1}$(0.2) = 11.3°, corresponding to an entrainment coefficient slightly lower than that reported in the literature[24] (= 0.24 ± 0.02).

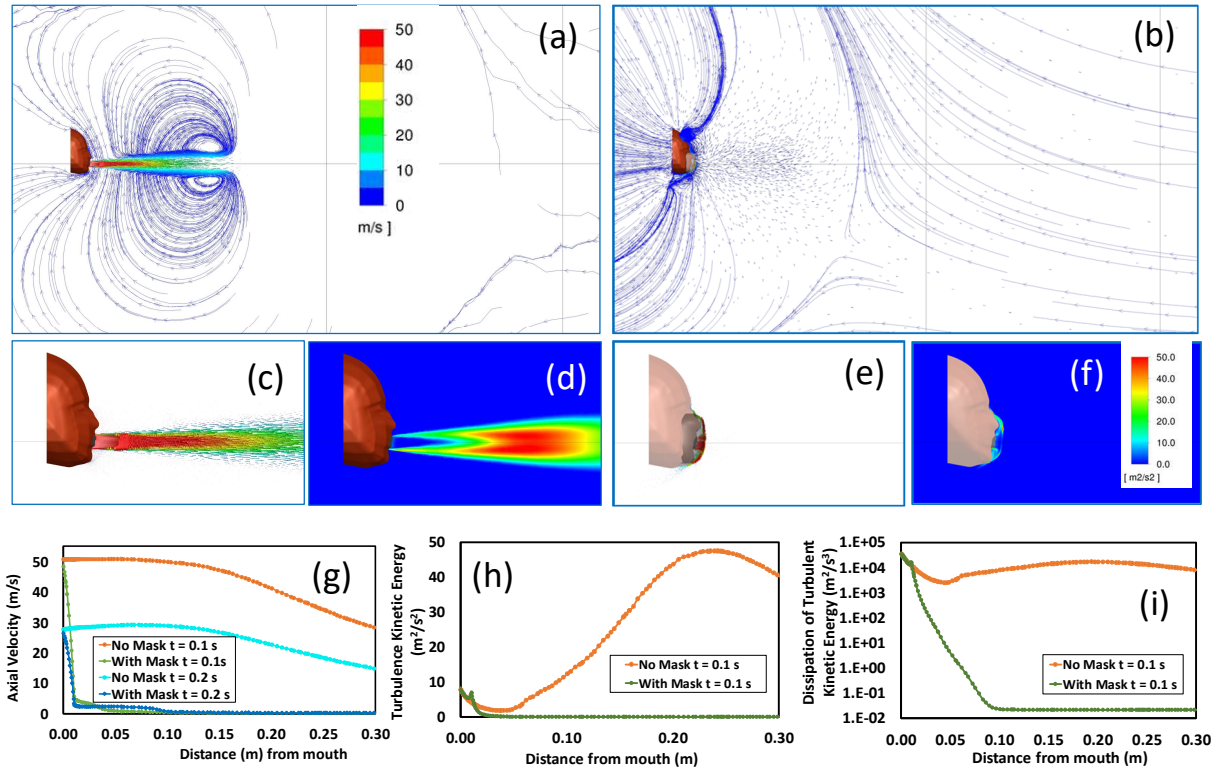

*Figure 1: Velocity vectors and ambient air streamlines showing reverse flows, close-ups of velocity streamline vectors and turbulent kinetic energy close to the face are presented at the symmetry plane passing through the face at t = 0.1 s after the respiratory event. Data is presented for the case without a mask (a), (c) and (d), and wearing a mask (b), (e) and (f), respectively. The colour scale for the magnitude of the velocity is shown as an inset in (a), while the scale for the turbulent kinetic energy is shown in (f). Axial velocity (g), turbulent kinetic energy (h) and dissipation (i) at the centreline passing through the mouth are shown for the respiratory event with and without the mask.*

There is a qualitative change in the airflow when a mask is worn (Figure 1 b, e). Here, the turbulent jet and strong recirculating flows are eliminated by the mask and about 12% of the air flow is diverted through the openings at the sides of the mask to create a qualitatively different flow around the face. We reiterate that the area of the openings (considering a uniform gap of 1 mm all around, between the face and the mask) represents only about 2% of the area of the mask. This leakage flow is similar to experimental reports[30] of flows using surgical masks (which are also not tightly fitted).

Correspondingly, we observe a drastic change in the spatial distribution of the turbulent kinetic energy when a mask is worn (compare Figure 1 d with 1 f, plotted through the symmetry plane at t = 0.1 s). Without a mask, a highly turbulent jet with large mean square velocity fluctuations propagates axially away from the face (Figure 1 d). This is virtually eliminated by the mask (Figure 1 f). At t = 0.1 s, the centerline velocity through the face decreases from ≈ 50 m/s to ≈ 40 m/s over 0.2 m. We note that the velocity reported here is exactly at the centreline. Therefore, these values are higher than the experimentally measured peak velocities of the jet, that are likely averaged around the centreline. At 0.2 s after the respiratory event, the centreline velocity immediately after the face is about 30 m/s, and decreases to about 20 m/s at 0.2 m. In contrast, the centreline velocities drop to less than 5 m/s within 0.02 m when a mask is worn (Figure 1 f).

Correspondingly, when no mask is worn, the centerline turbulent kinetic energy decreases from about 8 $m^2/s^2$ to 2 $m^2/s^2$ at 0.05 m from the face at t = 0.1 s, and then rises to 47 $m^2/s^2$ at a distance of 0.25m from the face as the entrained air forms a turbulent cloud (Figure 1 h). When a mask is worn, the turbulent kinetic energy rises near the mask due to the increase in mean square velocity fluctuations as the expelled jet impinges on the mask. However, due to the resistance to the flow presented by mask, it rapidly decreases immediately after the mask and approaches 0 $m^2/s^2$ at 0.04 m from the face. The dissipation of the turbulent flow field tracks the trend in the turbulent kinetic energy. Without a mask, it decreases by about 10-fold over 0.05 m (for t = 0.1 s) and then rises reaching a maximum at 0.2 m, while wearing a mask results in a rapid decrease by over 3 orders of magnitude over a distance of 0.05 m (Figure 1 i).

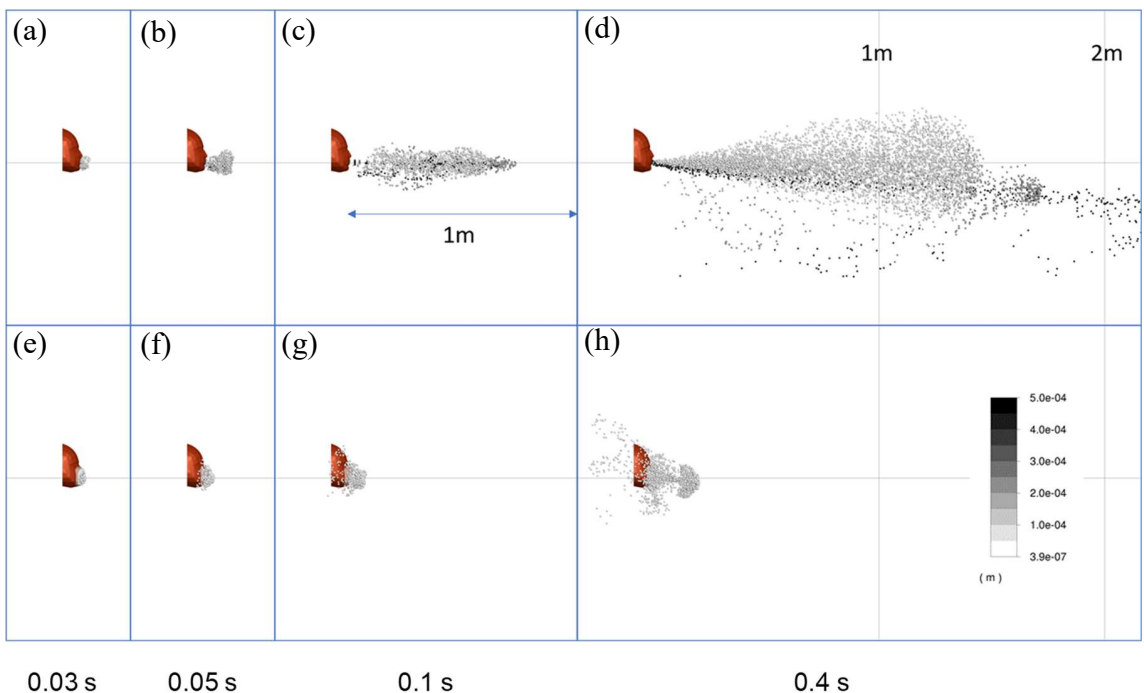

*Figure 2 Spreading of ejecta at different time instants after a respiratory event. Two sequences of images are shown at times t = 0.03 s (a, e), 0.05 s (b, f), 0.1 s (c, g) and 0.4 s (d, h). Top row shows images for the case without mask while the bottom row shows images for the case with mask. Lateral distances are as indicated in (c) and (d). The diameter of the droplets is represented by the scale bar at the bottom right.*

Wearing a mask has a significant impact on the spread of cough ejecta. We observe the time dependent trajectories of large and small droplets with time from the respiratory event (Figure 2). Without a mask (Figure 2, top panel), large drops are not convected by the flow and rapidly fall to the ground: drops > 200 μm fall within a lateral distance of 0.2 m, while drops > 125 μm extend to about 2 m (SI, Figure S6). In contrast to the large drops, smaller drops (< 25 μm in size) are convected by the turbulent cloud. They shrink in size as their water content is completely evaporated, and are transported to significant distances, as far as 5 m from the face (SI, Figure S7). We observe that the non-volatile content in these drops continues to stay suspended for as long as 60 s. Our data is consistent with the experimental literature.[25]

Wearing even a simple cotton mask restricts the spatial transport of droplets (Figure 2, bottom panel). Large droplets (> 4 μm) are trapped by the mask while smaller droplets are transported by the flows through the surface of the mask and through the openings on the sides. At t = 0.4 s, droplet ejecta is transported over less than 0.3 m (as compared to well over 2 m, without a mask). Thus, large droplets are trapped by the mask while the damping of the turbulent flow field by the mask leads to smaller droplets being transported only over relatively short distances. Flow through the openings around the mask convects small droplets along the face, in contrast to the case without a mask.

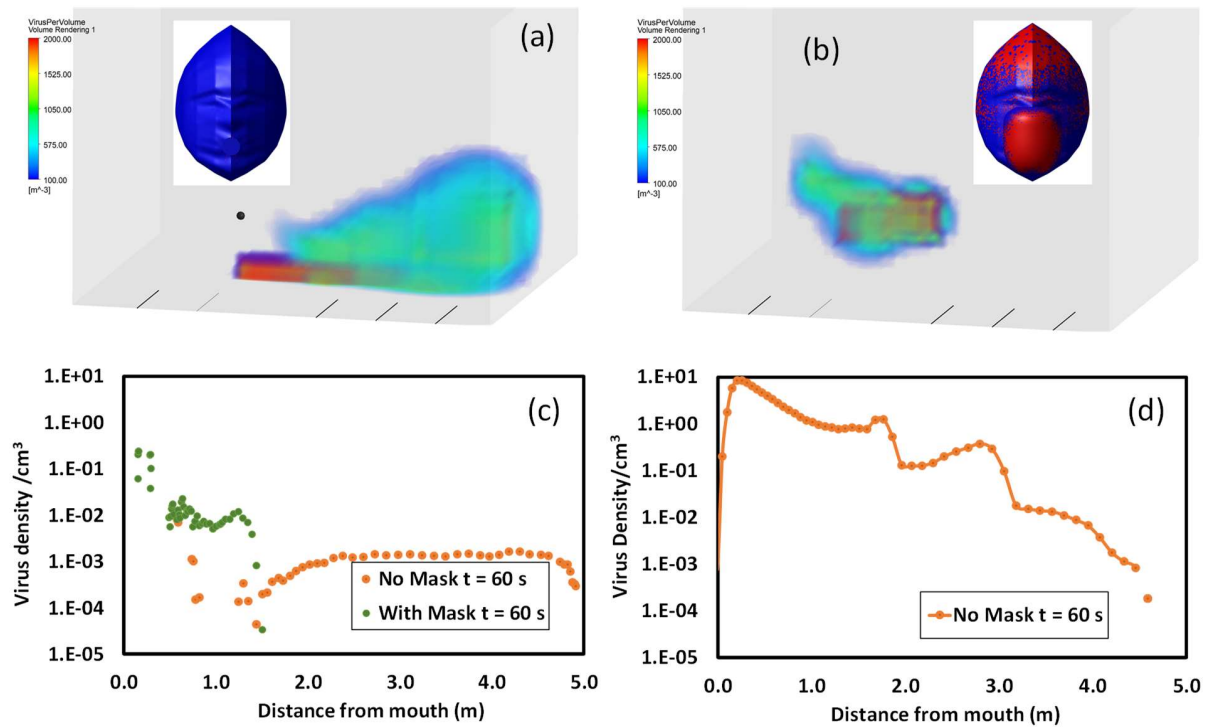

*Figure 3. Estimated density of virus particles ejected in the simulation box (a) without mask, and (b) with mask. The deposition on the face is shown in the inset of (a) while that on the face and mask is shown in (b). The density of suspended virus particles at t = 60 s, at the centreline through the mouth is shown in (c) while the deposited virus on the floor (at the centreline, at t = 60 s) is shown in (d).*

We estimate the potential viral concentrations suspended in the air and deposited on the floor due to propagation of droplets from the respiratory event of an infected person. Recent literature[37] indicates that the SARS-CoV2 load in throat swabs on patients within the first 5 days of the infection averages $6.76 \times 10^5$ RNA copies/mL. For sputum samples, an average of $7 \times 10^6$ RNA copies/mL was observed. Based on these, in these simulations we assume a representative viral concentration of $10^6$ particles/mL in the ejecta, to estimate the potential spatial dispersion of the virus. When an infected person not wearing a mask sneezes or coughs, virus particles in the large droplets rapidly drop to the floor. By t = 60 s, ≈ 37% of the potential viral load in the ejecta is deposited on the floor while ≈ 63% remains in the air. Most of the virus particles that are deposited on the floor are within 2 m from the person, with a maximum virus density ≈ 10 $cm^{-3}$ at about 0.2 m from the person (Figure 3 a, d). The suspended aerosolized virus particles form a low density cloud that extends from 2 to 5 m. At the centerline passing through the face, the suspended concentration ≈ $10^{-3}$ $cm^{-3}$ (Figure 3 a, c).

When a mask is worn, most of the virus-laden droplets (nearly 70%) are deposited on the mask. Flow through the mask surface and leakage flows from the openings around the mask result in generating a cloud, potentially conveying virus to a distance of about 1.5 m from the face. At the centerline passing through the face, this cloud has a density of $10^{-2}$ $cm^{-3}$ within about 1.5 m from the person (Figure 3 b, c). However, the suspended concentration drops significantly after 1.5 m from the person, and there is virtually no deposition of droplets on the ground (Figure 3 b, c). Thus, there is a clear qualitative difference in the distribution of virus particles when the infected person wears a mask. Without a mask, high concentrations of potentially virus-laden droplets are deposited on the floor within 2 m of the person and a dilute suspended cloud is observed over 2 – 5 m. In contrast, when a mask is worn, there is no deposit on the ground since most of the virus is deposited on the mask. Virus particles stay suspended within 1.5 m of the person, but this suspended concentration falls off sharply after that distance.

Our simulation results conclusively demonstrate that wearing even just a simple cotton mask has a dramatic influence on the air flow and spread of ejecta after a respiratory event. When a person not wearing a mask coughs or sneezes, the emanating jet sets up turbulent flows at distances of several meters from the person. While the large mucosal droplets fall to the floor within a distance of 2 m, the turbulent clouds continue to suspend aerosols at distances up to 5 m, for over a minute after the respiratory event. When an infected person coughs or sneezes, most of the virus deposits on the floor within a meter of the person. However, a dilute aerosol stays suspended, potentially carrying virus particles. In contrast, wearing a mask dissipates the turbulent flows passing through the mask, and diverts about 12% of the flow to the openings at the sides of the mask. The vast majority of the virus particles are retained on the mask and face. At t = 60 s, a cloud of virus particles (10-fold higher in density compared to the case without the mask) stays suspended within 1.5 m of the person. Our results strongly suggest that airborne transmission from patients (especially asymptomatic or presymptomatic patients) can be greatly reduced by wearing a simple cotton mask and maintaining strict physical distancing of 2 m.

# Supporting Information for:

# On the utility of cloth facemasks for controlling ejecta during respiratory events

Vivek Kumar,[a] Sravankumar Nallamothu,[a] Sourabh Shrivastava,[a] Harshrajsinh Jadeja,[a] Pravin Nakod,[a] Prem Andrade,[a] Pankaj Doshi,[b] Guruswamy Kumaraswamy[c]

[a] Ansys Software India Pvt. Ltd., Hinjewadi, Phase-1, Pune 411057, Maharashtra, India.
[b] B1-1510, Blue Ridge Township, Hinjewadi, Pune 411057, Maharashtra, India.
[c] Chemical Engineering Department, Indian Institute of Technology-Bombay, Mumbai 400071, Maharashtra, India.

## A. Simulation Methodology

The fluid phase is treated as a continuum by solving the Navier-Stokes equations, while the dispersed phase is solved by tracking droplets through the calculated flow field. The dispersed phase can exchange momentum, mass, and energy with the fluid phase.

### Discrete phase model

The dispersed phase is treated by the Lagrangian approach, where a large number of droplet parcels, representing a number of real droplets with the same properties, were traced through the flow field. By representing droplets by parcels, one can consider size distribution and simulate the measured liquid mass flow rate at the injection locations by a reasonable number of computational droplets. The trajectory of each droplet parcel is calculated by solving the equation of motion for a single droplet.

The droplets in the dispersed phase are modelled using the Discrete Phase Model (DPM). In this approach, a Lagrangian frame of reference is used to calculate the trajectories of a large number of droplets representing real droplets with the same properties, by integrating the forces acting on droplets. The droplets can exchange mass, momentum and energy with the fluid. The force balance on each droplet can be written as:

$$m_p \frac{du_p}{dt} = m_p \frac{(\vec{u} - \vec{u}_p)}{\tau_r} + m_p \frac{\vec{g}(\rho_p - \rho)}{\rho_p} + \vec{F} \qquad (1)$$

Where $m_p$ is the particle mass, $\vec{u}$ is the fluid velocity, $\vec{u}_p$ is the droplet velocity, ρ is the fluid density, ρp is the density of the droplet, $\vec{F}$ is an additional force, $m_p \frac{(\vec{u}-\vec{u}_p)}{\tau_r}$ is the drag force, and $\tau_r$ is the droplet relaxation time calculated by

$$\tau_r = \frac{\rho_p d_p^2}{18\mu} \frac{24}{C_d \, Re} \qquad (2)$$

Here, $\mu$ is the molecular viscosity of the fluid, and $d_p$ is the diameter of the droplet. The relative Reynolds number $Re$ is defined as

$$Re = \frac{\rho d_p |\vec{u}_p - \vec{u}|}{\mu} \quad (3)$$

The drag coefficient is calculated considering spherical particles. The dispersion of droplets due to turbulence in fluid phase is included using the stochastic tracking (random walk) model which includes the effect of instantaneous turbulent velocity fluctuations. For evaporating droplets, inert heating/cooling along with vaporization laws are applied. More information about the energy treatment of the DPM droplets and evaporation rate can be found in ANSYS Fluent 2020R1 Help Manual. Other fluid-droplet and droplet-droplet interactions are ignored in the study. The droplets are two-way coupled into the continuum fluid phase to make it possible for the droplets to influence the continuous fluid phase.

We model the effect of wearing a woven cloth face mask as follows:
The mask is included in the CFD model as a thin volume and modeled as porous media. For the current study the media is considered to be homogeneous and the resistances included via this media are considered to be isotropic in nature. Porous media are modeled by the addition of a momentum source term to the standard fluid flow equations. The source term is composed of two parts: a viscous loss term (Darcy, the first term on the right-hand side of Equation 4, and an inertial loss term (the second term on the right-hand side of Equation 4)

$$S_i = -\left( \sum_{j=1}^{3} D_{ij} \mu v_j + \sum_{j=1}^{3} C_{ij} \frac{1}{2} \rho |v| v_j \right) \quad (4)$$

where $S_i$ is the source term for the $i^{th}$ (x, y, or z) momentum equation, $\mu$ is the dynamic viscosity of fluid, v is the magnitude of the velocity and $D_{ij}$ and $C_{ij}$ are prescribed matrices. This momentum sink contributes to the pressure gradient in the porous cell, creating a pressure drop that is proportional to the fluid velocity (or velocity squared) in the cell.
To recover the case of simple homogeneous porous media

$$S_i = -\left( \frac{\mu}{\alpha} v_i + \sum_{j=1}^{3} C_2 \frac{1}{2} \rho |v| v_j \right) \quad (5)$$

where $\alpha$ is the permeability and $C_2$ is the inertial resistance factor, obtained by specifying **D** and **C** as diagonal matrices with diagonal values of $1/\alpha$ and $C_2$, respectively (and zero for the other elements).

We consider a mask prepared from cotton cloth. Details of the cloth used are taken from the thesis[1] of Saldaeva. Details of the construction of the fabric as provided in the thesis are given in Table S1.

| Mask | Looseness factor | Fiber radius | Yarn fiber volume fraction | Thickness | Yarn spacing warp | Yarn spacing weft | Yarn width warp | Yarn width weft |
|---|---|---|---|---|---|---|---|---|
| 100% cotton | 0.32 | 4.3 μm | 0.56 | 323 μm | 470 μm | 410 μm | 405 μm | 279 μm |

*Table S1: Mask Fabric Specification (from the thesis of Saldaeva[1])*

Pressure drop versus velocity data obtained from the thesis and shown below in Figure S1 can be fitted to obtain the Darcy (proportional to v) and non-Darcy inertial (proportional to $v^2$) components of the resistance to flow.

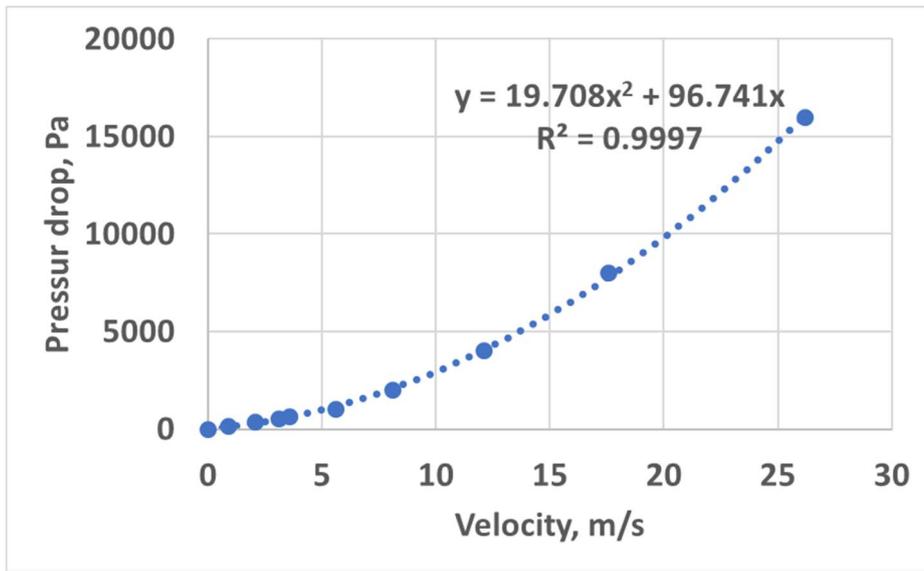

*Figure S1: Pressure drop vs Velocity for a typical 100% cotton fabrics. A quadratic trendline is fitted to get the coefficients for Porous Media Modeling*

For the CFD method validation, a model similar to the one described in Aliabadi et. al.[2] is developed. A geometry representing 1m x 2m x 2m domain with a small duct of 0.2m x 0.2m placed opposite side of the inlet plane. The exact location of the duct was not mentioned in the reference. This could affect the flow field especially towards the duct. For simplicity, half of the domain is simulated by creating a symmetry boundary. A 2 $cm^2$ area is created to represents mouth area. The computational domain is discretized into 425K polyhedral cells. A snapshot of the computational domain marked with boundary conditions and mesh on symmetry plane is shown in Figure S2.

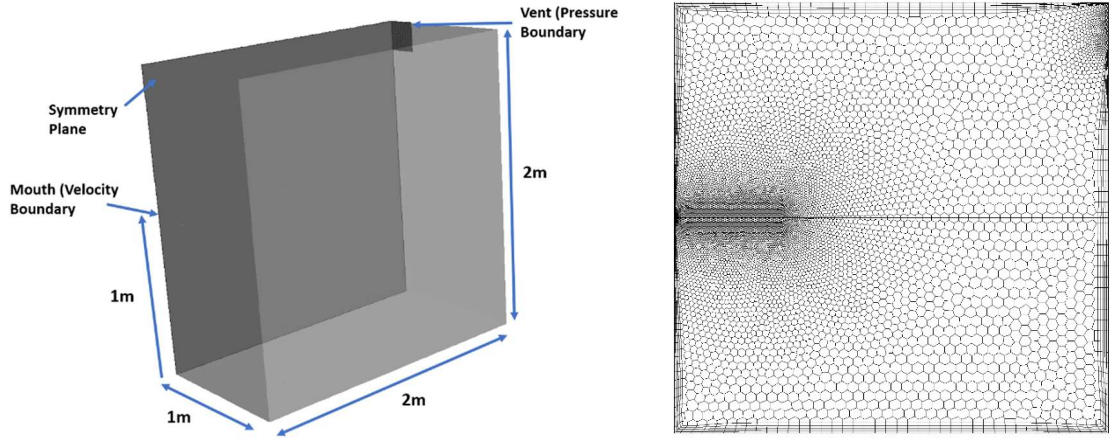

*Figure S2: Computational domain & mesh considered for baseline case.*

For the jet coming from the mouth, a time dependent velocity profile is applied with peak velocity time (PVT) of 0.1 sec as shown in Figure S3. Cough peak flowrate (CPFR) is derived using the formulation reported by Gupta et al.[3] To simulate varied human expirations, various cough expired volume (CEV) values are used corresponding to very weak, medium and very strong expirations.

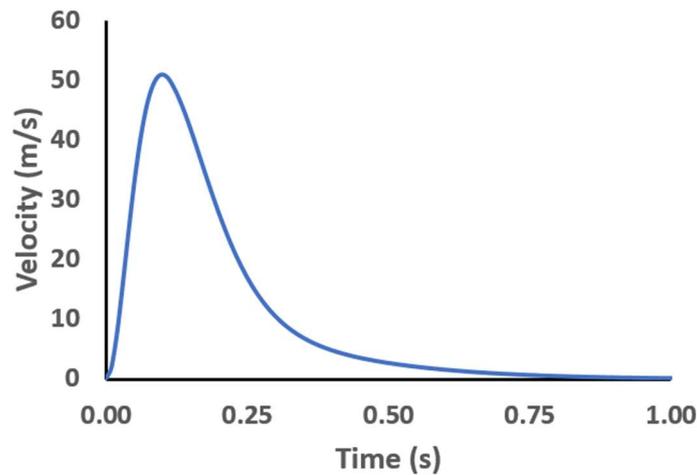

*Figure S3: Time dependent velocity profile applied at the mouth*

To simulate the spray of droplets, a Rosin-Rammler distribution method is used for the injection. 94% of injected droplet mass fraction represents evaporating sprays and the remaining 6% represents the non-volatile matter. A total 6.1 mg of droplets are injected using 10 injections with equal distribution among the injections. The initial droplet positions are staggered over 0.01 m. Conditions used for the baseline simulation are shown in Table S2.

| Ambient Temperature | 23 °C |
|---|---|
| Ambient Relative Humidity | 20% |
| Ambient density | Air density @ 23 °C |
| Ambient velocity | 0 m/sec |
| Evaporation Model | Diffusion controlled |
| Droplet distribution | Logarithmic RR with min =1 µm, max = 500 µm, mean = 10 µm |

*Table S2: Simulation conditions used for baseline case*

Porous media properties of the cotton mask such as the permeability and porous jump coefficients are needed for considering the effect of mask on fluid flow. Saldaeva[1] measured through-thickness air permeability of different textile materials and reported the permeability values and non-Darcy flow coefficients. From this study, porous media properties of 100% plain cotton fabric have been utilized for providing inputs to the porous jump boundary condition representing the cloth mask.

A user defined function (UDF) has been used to filter the droplets from the mist generated by the respiratory event. We model filtration by cloth with reference to reported experimental literature. Guyton et. al.[4] measure the filtration efficiency of a single layer of fabric typical of bath towels, cotton shirts, handkerchiefs, etc. and report the penetration efficiency of 2 µm particles. Rengasamy et. al.[5] report the penetration efficiency of cotton cloth for a range of particle sizes (up to 1 µm in size). We use a conservative estimate for the penetration efficiency of the cotton mask used in this work and implement a simplified filtering mechanism such that all droplets above 4 µm diameter are filtered out while droplets below 4 µm are allowed to pass through the mask. This boundary condition is applied when the droplets hit the surface of the mask.

## Validation of the CFD model

The baseline case is created to validate the methods used simulate near field cough, particle dispersion, heat and mass transfer in a still environment. The results from the baseline CFD simulation are verified against results from Aliabadi et. al.[2] In figure S4 (a), average droplet diameter in each of the diameter bins are plotted against time. Figure S4(b) shows the vertical penetration of the droplet plumes in the gravity direction. In the current work, penetration is computed as the location where 98% of the total mass of the droplet is contained. The exact approach taken by Aliabadi et. al.[2] to compute penetration length is not known. Some differences may be expected because of this uncertainty as wells as the uncertainty about the exact location of the duct.

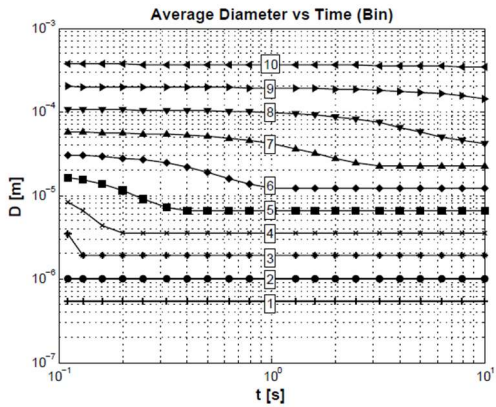 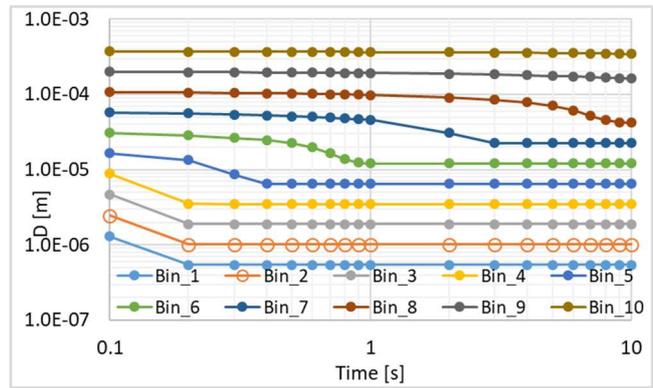

a) Binned Droplet Diameters

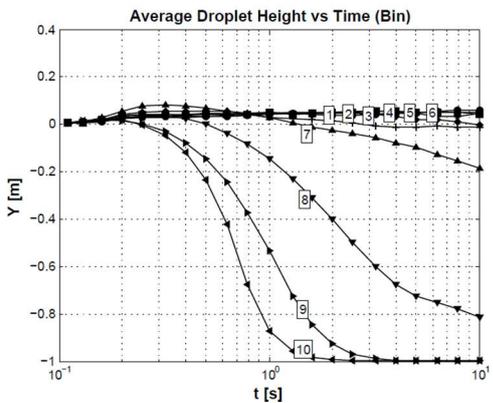 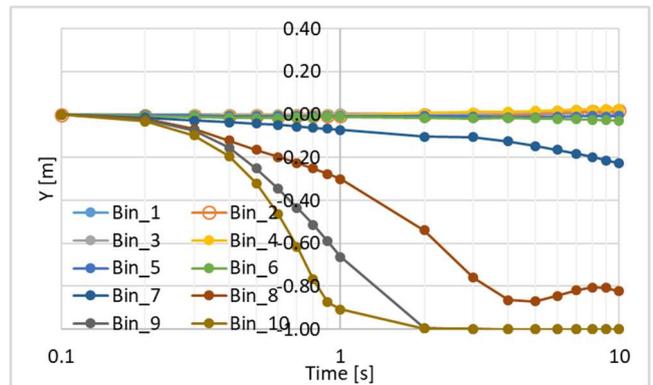

b) Vertical Penetration

*Figure S4: Comparison of axial and vertical penetration of binned droplets compared with Aliabadi et. al.[2]*

## Additional Information and Results

We employ a geometry representing a 5m x 6m x 6m domain with a small duct of 0.2m x 0.2m placed opposite side of the inlet plane. Half of the computational domain is simulated by placing a symmetry boundary condition in the middle. The computational domain is discretized into 1.8 million elements with a mix of Polyhedral and hexahedral elements using ANSYS Fluent 2020R1 mesh creation tools as shown in Figure S5.

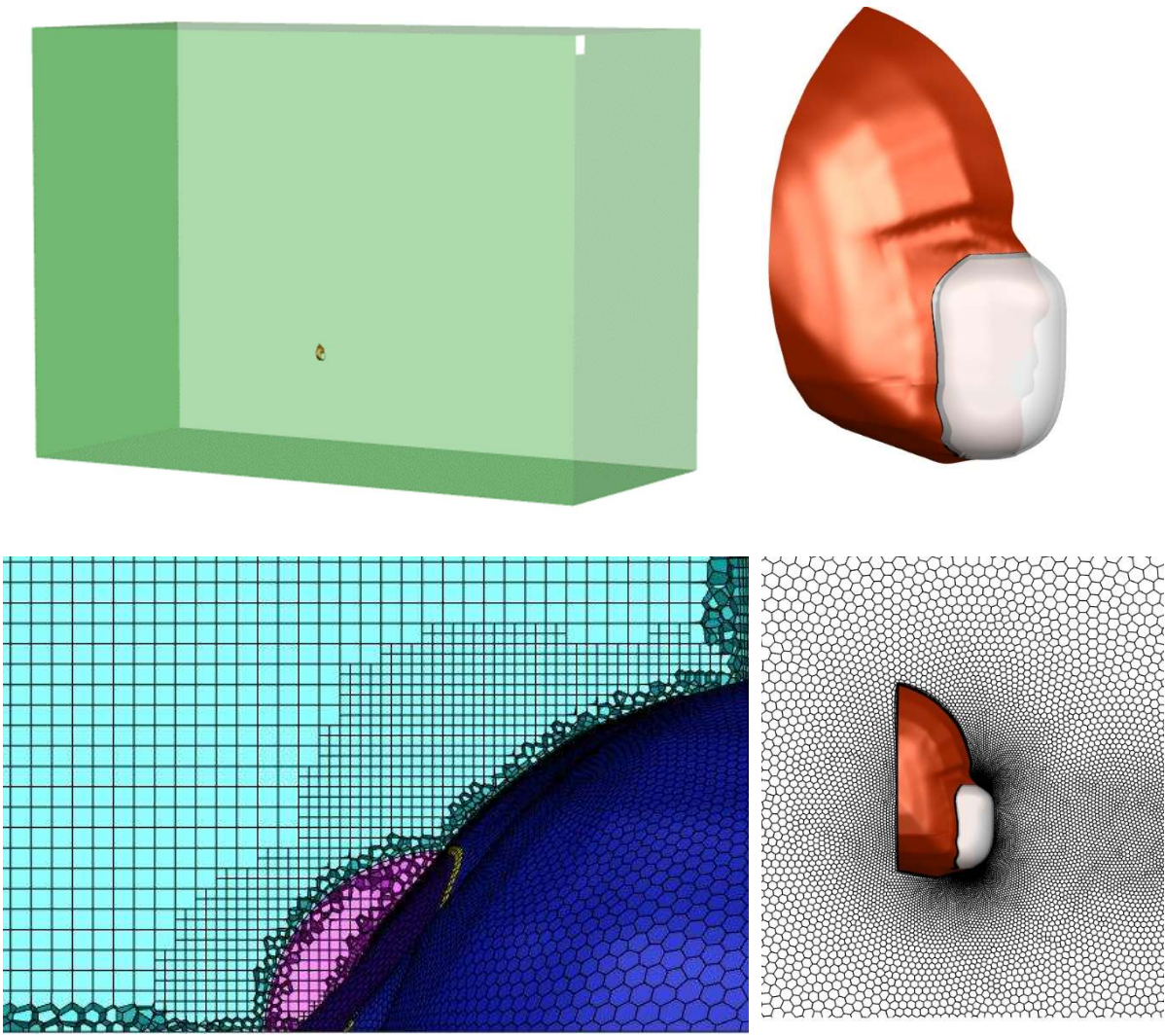

*Figure S5: Computational domain and mesh used for cough simulation with face mask*

For evaluation of the impact of a face mask, a human face is included in the domain at a height of 1 m from the ground.

The conditions used for the simulation presented in this paper are shown in Table S3.

| Ambient Temperature | 35 °C |
|---|---|
| Ambient Relative Humidity | 60% |
| Ambient density | Air density @ 35 °C |
| Ambient velocity | 0 m/sec |
| Evaporation Model | Diffusion controlled |
| Droplet distribution | Logarithmic Rosin Rammler<br>Min size = 1 μm, Max size = 500 μm<br>Mean size = 10 μm, Spread = 0.1 |
| Ejecta composition | 94% volatile and 6% inert droplets |

*Table S3: Simulation conditions used for the work presented here.*

B. Additional simulation results, as referred to in the main manuscript are presented here.

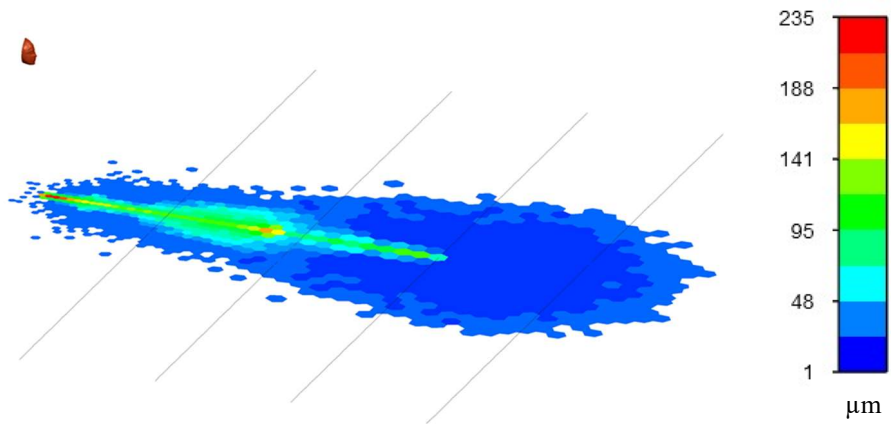

*Figure S7: Average Diameter of the droplet at the time of hitting the floor. Red zone near the individual location is indicative of heavier droplets settling down. The lines represent lateral distances of 1 m.*

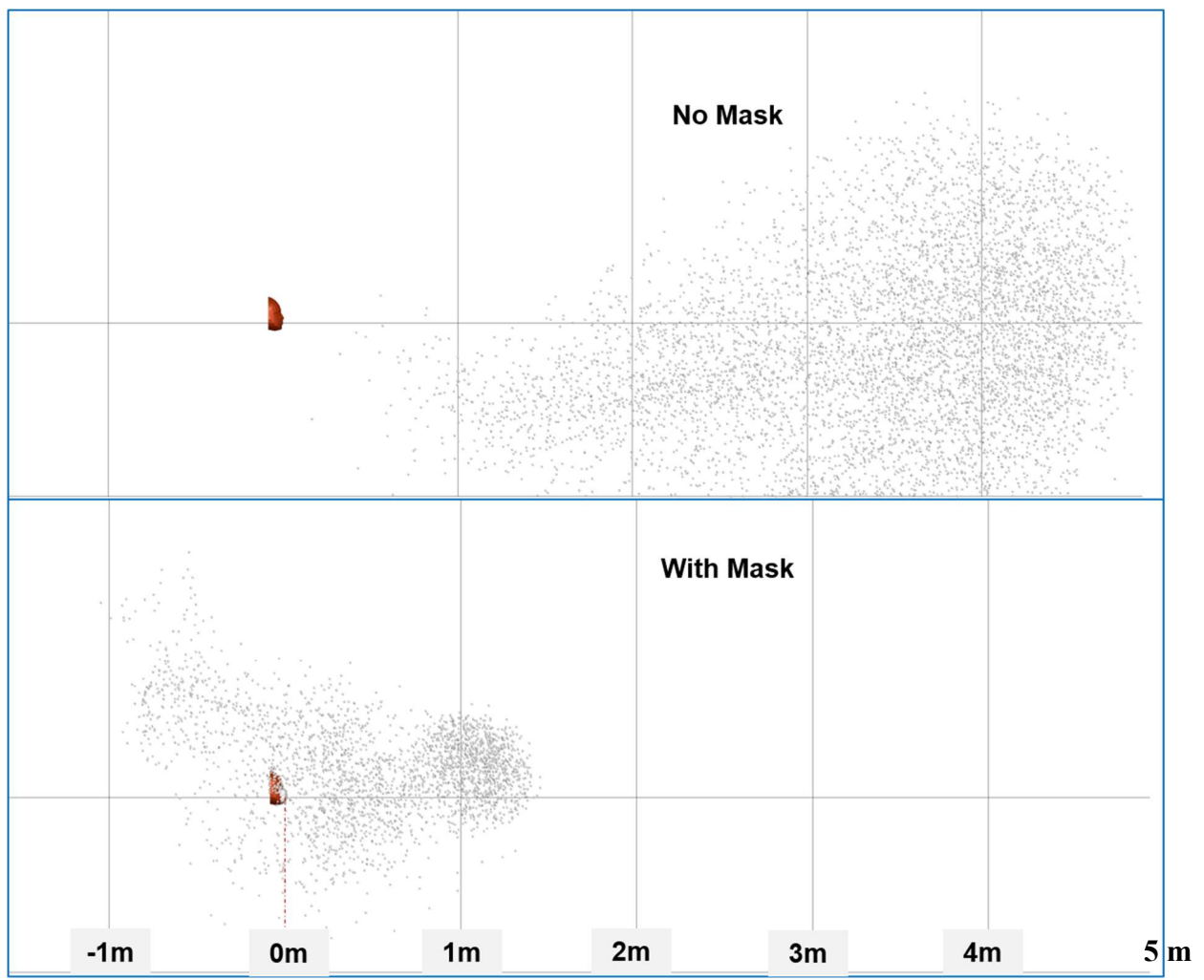

*Figure S6: Droplets of the diffused aerosol location after 1 minute. Without mask the droplet plume has travelled nearly 5 meters. With mask, the droplet plume is concentrated within 1.25 meters from the individual.*